\documentclass[journal, 10pt]{IEEEtran}
\usepackage{cite}
\usepackage{graphicx}
\usepackage{amsmath}
\usepackage{amssymb}
\usepackage{amsthm}
\usepackage{amsfonts}
\usepackage{algorithm}
\usepackage[noend]{algpseudocode}
\usepackage{array}
\usepackage{color}
\usepackage{enumerate}
\usepackage{subfigure}

\makeatletter

\usepackage{algorithm,algpseudocode,float}
\usepackage{lipsum}



\begin{document}

\title{Heterogeneous Transformer: A Scale Adaptable Neural Network Architecture for
Device Activity Detection}


\author{Yang~Li,~
        Zhilin~Chen,~
        Yunqi~Wang,~
        Chenyang~Yang,
        ~and~Yik-Chung~Wu

\thanks{Y. Li is with Shenzhen Research Institute of Big Data,
Shenzhen, 518172, China (e-mail: liyang@sribd.cn).}
\thanks{Z. Chen is with The Edward S. Rogers Sr. Department of Electrical
and Computer Engineering, University of Toronto, Toronto, ON M5S 3G4, Canada (e-mail: zchen@comm.utoronto.ca).}
\thanks{Y. Wang and Y.-C. Wu are
with the Department of Electrical and Electronic Engineering, The
University of Hong Kong, Hong Kong (e-mail: \{yunqi9@connect, ycwu@eee\}.hku.hk).}
\thanks{C. Yang is with the School of Electronics and Information
Engineering, Beihang University, Beijing, 100191, China (e-mail: cyyang@buaa.edu.cn).}}

\maketitle
\begin{abstract}
To support the modern machine-type communications,
a crucial task during the random access phase
is device activity detection, which is
to detect the active devices from a large number of potential devices based on the received signal
at the access point.
By utilizing the statistical properties of the channel,
state-of-the-art covariance based methods have been demonstrated to achieve better
activity detection performance than compressed sensing based methods.
However, covariance based methods require to
solve a high dimensional nonconvex optimization problem by updating
the estimate of the activity status of each device sequentially. Since the number of updates is proportional to the device number, the computational complexity and delay make the iterative updates
difficult for real-time implementation especially when the device number scales up.
Inspired by the success of deep learning for real-time inference,
this paper proposes a learning based method with a customized heterogeneous transformer architecture for device activity detection.
By adopting an attention mechanism in the architecture design,
the proposed method is able to extract the relevance between device pilots and received signal,
is permutation equivariant with respect to devices, and is scale adaptable to different numbers of devices.
Simulation results demonstrate that the proposed method achieves better activity detection performance
with much shorter computation time than state-of-the-art covariance approach, and
generalizes well to different numbers of devices, BS-antennas, and different signal-to-noise ratios.
\end{abstract}



\section{Introduction}
To meet the dramatically increasing demand for wireless connectivity of Internet-of-Things (IoT),
machine-type communications (MTC) have been recognized as a new paradigm in the fifth-generation and beyond wireless systems.
Different from the traditional human-to-human communications, MTC scenarios commonly involve
a large number of IoT devices connecting to the network, but only a small portion of the devices are active at any given time due to the sporadic traffics \cite{Bockelmann2016,Dawy2017,LiuLiang2018}.
In activity detection, each device is assigned a unique pilot sequence
and the base station (BS) detects which pilot sequences are received in the random access phase \cite{LiuLiang2018-1,LiuLiang2018-2}.
However, the pilot sequences for device activity detection have to be nonorthogonal,
\textcolor{black}{due to the large number of devices but limited coherence time}.
The nonorthogonality of the pilot sequences inevitably
induces interference among different devices, and hence complicates the task of device activity detection in MTC.

To tackle the problem of device activity detection with nonorthogonal pilot sequences,
two major approaches have been proposed in the literature. The first approach
identifies the active devices
through joint device activity detection and channel estimation using compressed sensing based methods \cite{Chen2018,SunZhuo2019,Senel2018,Jiang2020,Utkovski2017,Chen2019,Ke2021,XuXiao2015,Ahn2019,Liu2018,He2018,Li2019,Shao2020}. Specifically, \cite{Chen2018,SunZhuo2019} proposed
approximate message passing (AMP) based algorithms to jointly
recover the device activity and the instantaneous channel state information.
Furthermore, AMP was extended to include data detection \cite{Senel2018,Jiang2020}
and to multi-cell systems \cite{Utkovski2017,Chen2019,Ke2021}, respectively.
In addition to AMP, other compressed sensing based methods, such as
Bayesian sparse recovery \cite{XuXiao2015,Ahn2019}
and regularization based sparse optimization \cite{Liu2018,He2018,Li2019,Shao2020}
have also been investigated for joint device activity detection and channel estimation.

Different from the compressed sensing based methods,
the second approach utilizes the
statistical properties of the channel
without the need of
estimating the instantaneous channel state information.
This approach is referred to as the covariance based methods,
since they are based on the sample covariance matrix of the received signal \cite{Haghighatshoar2018,Shao2020-2,Jiang2020-2,Chen2021,Ganesan2021}.
The covariance based methods have recently drawn a lot of attention
due to the superiority of activity detection performance. In particular, the analytical results in \cite{Fengler2021,Chen2020} show that
the required pilot sequence length of the covariance based methods
for reliable activity detection is much shorter than that of the compressed sensing based methods.

While the covariance based methods outperform
the compressed sensing based methods due to
the advantage of utilizing the statistical properties of the channel,
the covariance approach requires to
solve a high dimensional nonconvex optimization problem \cite{Haghighatshoar2018,Shao2020-2,Jiang2020-2,Chen2021,Ganesan2021}, where the estimate of the
activity status of each device is updated sequentially using the coordinate descent method.
The sequential nature of the coordinate descent method implies that the number of updates is proportional to the total number
of devices. Consequently, the resulting computational complexity and delay make
it unsuitable for real-time implementation, especially when the device number is very large.

Recently, deep learning has been exploited to avoid the high computational cost caused by
iterative algorithms \cite{SunHR2018,Zhu20,Shen2021,Guo2021}.
Instead of solving each optimization problem case-by-case, deep learning utilizes neural networks to represent a mapping function from many problem instances to the corresponding solutions based on a large number of training samples.
Once the mapping function is obtained, the neural network can infer
the solution of any new problem in a real-time manner.
Moreover, thanks to the universal approximation property of neural networks \cite{Hornik89},
deep learning also has the opportunity to learn a better solution than the conventional model-based methods for complex problems \cite{Eisen2020,Sohrabi21,Jiang2021,Li21}.

The potentials of deep learning
in computational efficiency and solution quality
motivate us to study a learning based method for
tackling the high dimensional nonconvex problem in device activity detection.
For this purpose, we interpret the activity detection as
a classification problem and
design a customized neural network architecture
for representing the mapping function from the received signal and device pilots
to the corresponding activity status.
While generic multi-layer perceptrons (MLPs) have been widely used for function representations,
they are not tailored to the activity detection problem
due to the lack of some key properties. In particular,
\begin{enumerate}[$\bullet$]
\item
To detect the device activities, the BS should perceive which device pilots are received from the received signal.
Therefore, it is beneficial to incorporate a computation mechanism into the neural network
for evaluating the relevance between the received signal and device pilots.
However, generic MLPs do not generate such an attribute.
\item
The device activity detection has an inherent permutation equivariance property
with respect to devices. To be specific, if the indices of any two devices are exchanged,
the neural network should output a corresponding permutation.
Incorporating permutation equivariance into the neural network architecture can reduce the
parameter space and also avoid a large number of unnecessary permuted training samples \cite{Shen2021,Guo2021,Eisen2020}.
Unfortunately, the architecture of generic MLPs cannot guarantee the permutation equivariance
property.
\item
As the device number scales up,
it is highly expected that
the neural network is generalizable to larger numbers of devices
than the setting in the training procedure.
Nevertheless, generic MLPs are designed for a
pre-defined problem size with fixed input and output dimensions,
and thus the well-trained MLPs are no longer applicable to a different number of devices.
\textcolor{black}{
Recent} works \textcolor{black}{have} applied deep learning for \textcolor{black}{joint activity detection and pilot design
\cite{Cui2021} and for joint activity detection and channel estimation \cite{Shi2021}}, respectively.
However, using the neural networks of \cite{Cui2021,Shi2021},
the device pilots can only be either optimized for a pre-defined device number \cite{Cui2021}
or fixed as a given matrix \cite{Shi2021}, and thus the well-trained neural networks can neither generalize to
a different device number nor \textcolor{black}{to} a different set of device pilots.
\end{enumerate}

To incorporate the properties mentioned above into the neural network,
this paper proposes a heterogeneous transformer architecture,
which is inspired by the recent successes
of the transformer model in natural language processing (NLP) \cite{Vaswani2017}.
In particular, transformer is built on an attention mechanism, which can extract the relevance among different words within a sentence. Based on the relevance extraction, transformer can decide which parts of the source sentence to pay attention to. We observe that
the relevance extraction
is appealing to the activity detection problem, since
the BS should perceive which device pilots contribute to the received signal
by evaluating the relevance between the received signal and
device pilots.

Yet different from the NLP tasks where
different words belong to the same class of features, the received signal and device pilots
in the activity detection problem have different physical meanings, and hence
should be processed differently.
This observation motivates us to design a heterogeneous transformer architecture.
Specifically,
instead of using a single set of parameters to process all the inputs as in the standard transformer,
we use two different sets of parameters, one
to process the representations
corresponding to the device pilots,
and the other set for the received signal.

The overall deep neural network consists
of an initial embedding layer, multiple heterogeneous transformer encoding layers, and a decoding layer.
The initial embedding layer takes the received signal and
device pilots as the inputs and
produces the initial embeddings. The initial embeddings are further processed through the encoding layers, where the heterogeneous attention mechanism is applied to extract the relevance among the received signal and device pilots. Finally, the decoding layer decides the activity status of each device
based on the extracted relevance.

The main contributions of this work are summarized as follows.
\begin{enumerate}[1)]
\item
We provide a novel perspective on how device activity detection
can be formulated as a classification problem with the received signal and device pilots as inputs.
By constructing a training data set of received signals and devices pilots
with ground-truth labels, we propose a deep learning approach to mimic the optimal mapping function from the received signal and device pilots to device activities, which
can potentially achieve better detection performance than state-of-the-art covariance approach. Moreover, instead of iteratively solving an optimization problem case-by-case,
the proposed learning based method can infer the solution of any new problem in a real-time manner due to the computationally cheap inference.

\item
We further propose a customized
heterogeneous transformer architecture based on the attention mechanism,
which learns the device activities from the relevance between the received signal and device pilots.
Moreover, by sharing the parameters for producing the representations of different device pilots, the proposed heterogeneous transformer is permutation equivariant with respect to devices,
and the dimensions of parameters that
require to be optimized during the training procedure
are independent of the number of devices.
This scale adaptability
makes the proposed architecture generalizable to different numbers of devices.

\item
Simulation results show that the proposed learning based method using heterogeneous transformer
achieves better activity detection performance with much shorter computation time than state-of-the-art covariance approach. The proposed method
also generalizes well to different numbers of devices, BS-antennas, and different
signal-to-noise ratios (SNRs).
\end{enumerate}

The remainder of this paper is organized as follows. System model and existing approaches
are introduced in Section II.
A novel deep learning perspective on device activity detection is proposed
in Section III. A heterogeneous transformer architecture is designed in Section IV.
Simulation results are provided in Section V. Finally,
Section VI concludes the paper.

Throughout this paper, scalars, vectors, and matrices are denoted by lower-case letters, lower-case bold letters, and upper-case bold letters, respectively. The real and complex domains are denoted by $\mathbb{R}$ and $\mathbb{C}$, respectively.
We denote the transpose, conjugate transpose, inverse,
real part, and imaginary part
of a vector/matrix by $(\cdot)^T$, $(\cdot)^H$, $(\cdot)^{-1}$, $\Re(\cdot)$, and $\Im(\cdot)$, respectively.
The $N\times N$ identity matrix and the length-$N$ all-one vector are
denoted as $\mathbf{I}_N$ and $\mathbf{1}_N$, respectively.
The trace, determinant, and the column vectorization of a matrix are represented as
$\mathrm{Tr}(\cdot)$, $\vert\cdot\vert$, and $\text{vec}(\cdot)$,
respectively.
The notation
$\odot$ denotes the element-wise product,
$\mathbb{I}(\cdot)$ denotes the indicator function,
$\text{ReLu}(\cdot)$ denotes the function $\max(\cdot,0)$,
and $\mathcal{CN}\left(\cdot, \cdot\right)$
denotes the complex Gaussian distribution.

\section{System Model and Existing Approaches}
\subsection{System Model}
Consider an uplink multiple-input multiple-output (MIMO) system with one $M$-antenna BS
and $N$ single-antenna IoT devices.
We adopt a block-fading channel model, where
the channel from each device to the BS
remains unchanged within each coherence block.
Let $\sqrt{g_n}\mathbf{h}_n$ denote the channel from the
$n$-th device to the BS, where $\sqrt{g_n}$ and $\mathbf{h}_n\in\mathbb{C}^{M}$
are the large-scale and small-scale Rayleigh
fading components, respectively.
Due to the sporadic traffics of MTC, only $K\ll N$ devices are active in each
coherence block. If the $n$-th device is active, we denote the activity status as $a_n=1$
(otherwise, $a_n=0$).

To detect the activities of the IoT devices at the BS, we assign each device a unique pilot sequence
$\mathbf{s}_n\in\mathbb{C}^{L_\text{p}}$, where $L_\text{p}$
is the length of the pilot sequence.
Device $n$ transmits the pilot sequence $\mathbf{s}_n$ with transmit power $p_n$
if it is active. Assuming that the transmission from different devices are synchronous,
we can model the received signal at the BS as
\begin{equation}\label{received signal}
\mathbf{Y} =
\sum_{n=1}^{N} \mathbf{s}_n\sqrt{p_ng_n}a_n
\mathbf{h}_n^T+\mathbf{W}
=\mathbf{S}\mathbf{G}^{\frac{1}{2}}\mathbf{A}
\mathbf{H}+\mathbf{W},
\end{equation}
where
$\mathbf{S}\triangleq\left[\mathbf{s}_1, \ldots, \mathbf{s}_N\right]\in\mathbb{C}^{L_\text{p}\times N}$,
$\mathbf{G}\triangleq\text{diag}\left\{p_1g_1, \ldots, p_Ng_N\right\}$,
$\mathbf{A}\triangleq\text{diag}\left\{a_1, \ldots, a_N\right\}$,
$\mathbf{H}\triangleq\left[\mathbf{h}_1, \ldots, \mathbf{h}_N\right]^T\in\mathbb{C}^{N\times M}$,
and $\mathbf{W}\in\mathbb{C}^{L_\text{p}\times M}$ is the Gaussian noise at the BS.

This paper aims to detect the activity status $\{a_n\}_{n=1}^N$
based on the received signal $\mathbf{Y}$
at the BS. Since the IoT devices are stationary in many practical deployment scenarios,
the large-scale fading components can be obtained in advance
and hence assumed to be known \cite{Jiang2020-2,Chen2021,Ganesan2021}.
In order to reduce the channel gain variations among different devices,
the transmit power of each device can be controlled based on the large-scale \textcolor{black}{channel gain} \cite{Senel2018}.
This is especially beneficial to the devices with relatively weak channel gains.

\subsection{Existing Approaches}
Existing approaches for device activity detection can be roughly divided into two categories:
compressed sensing based methods and covariance based methods.

\emph{1) Compressed Sensing Based Methods:}
Due to the sporadic traffics of MTC,
the activity status and
the instantaneous small-scale fading channels
can be jointly estimated by solving a compressed sensing problem.
Specifically, by denoting $\mathbf{B}\triangleq\mathbf{S}\mathbf{G}^{\frac{1}{2}}\in\mathbb{C}^{L_\text{p}\times N}$ and $\mathbf{X}\triangleq\mathbf{A}\mathbf{H}\in\mathbb{C}^{N\times M}$,
according to \eqref{received signal},
the activity status
can be obtained by recovering the row-sparse matrix $\mathbf{X}$
from $\mathbf{Y} = \mathbf{B}\mathbf{X}+\mathbf{W}$. However, since a large amount of
instantaneous channel state information requires to be estimated simultaneously,
the activity detection performance of compressed sensing based methods cannot compete with that of
covariance based methods.

\emph{2) Covariance Based Methods:}
When the BS is equipped with multiple antennas,
by utilizing the statistical properties of the channel,
covariance based methods can estimate the activity status
without estimating the instantaneous channels.
Specifically, the covariance approach treats
the small-scale fading channel matrix $\mathbf{H}$ and the noise matrix $\mathbf{W}$ as complex Gaussian random variables.
Each column of $\mathbf{H}$ and $\mathbf{W}$ are assumed to
follow independent and identically distributed
(i.i.d.) $\mathcal{CN}(\mathbf{0}, \mathbf{I}_N)$ and $\mathcal{CN}(\mathbf{0}, \sigma^2\mathbf{I}_{L_\text{p}})$, where $\sigma^2$ is the noise variance.
Let $\mathbf{y}_m$ denote the $m$-th column of the received signal $\mathbf{Y}$,
which follows i.i.d. $\mathcal{CN}\left(\mathbf{0}, \boldsymbol{\Sigma}\right)$ with $\boldsymbol{\Sigma}
=\mathbb{E}\left[\mathbf{y}_m\mathbf{y}_m^H\right]
=\mathbf{S}\mathbf{G}\mathbf{A}\mathbf{S}^H+\sigma^2\mathbf{I}_{L_\text{p}}$.
Consequently, the activity status $\{a_n\}_{n=1}^N$ can be detected
by maximizing the likelihood function\textcolor{black}{\cite{Jiang2020-2,Chen2021,Ganesan2021}}
\begin{eqnarray}\label{likelihood}
p(\mathbf{Y};\{a_n\}_{n=1}^N)
&=&\prod_{m=1}^{M}p(\mathbf{y}_m;\{a_n\}_{n=1}^N)\nonumber\\
&=&\frac{1}{\vert\pi\boldsymbol{\Sigma}\vert^M}\exp\left(-\text{Tr}\left(\boldsymbol{\Sigma}^{-1}\mathbf{Y}\mathbf{Y}^H\right)\right),
\end{eqnarray}
which is equivalent to the following combinatorial optimization problem
\begin{subequations}\label{opt0}
\begin{equation}\label{opt0_cost}
\min_{\{a_n\}_{n=1}^N}~~
\log\vert\boldsymbol{\Sigma}\vert+\frac{1}{M}\text{Tr}\left(\boldsymbol{\Sigma}^{-1}\mathbf{Y}\mathbf{Y}^H\right),
\end{equation}
\begin{equation}\label{opt0_constraint}
\text{s.t.}\
~~a_n\in\left\{0,1\right\},~~\forall n=1,\ldots,N.
\end{equation}
\end{subequations}
The covariance approach first relaxes the binary constraint \eqref{opt0_constraint}
as $a_n\in\left[0,1\right]$, and then applies the coordinate descent method that
iteratively updates each $a_n$ for solving the relaxed problem.
However, since the coordinate descent method requires to update each $a_n$
sequentially, the total iteration number
is proportional to the number of devices $N$, which induces
tremendous computational complexity and delay,
especially when the device number is massive.
Moreover, due to the non-convexity of the cost function \eqref{opt0_cost},
the covariance approach can only obtain a stationary point of the relaxed problem.

Different from the conventional optimization based algorithms which involve a lot of iterations,
deep learning based methods can \textcolor{black}{provide} a real-time solution by computationally cheap operations
\cite{SunHR2018,Zhu20,Shen2021,Guo2021}.
In the following two sections, we propose a deep learning based method for device activity detection.
The proposed method consists of
interpreting device activity detection as a classification problem (Section~III),
and a customized neural network architecture (Section~IV).

\section{A Novel Deep Learning Perspective}
\subsection{Device Activity Detection as Classification Problem}
In this paper,
we strive to learn the the activity status $\{a_n\}_{n=1}^N$ without estimating the instantaneous channel $\mathbf{H}$.
However, instead of learning to optimize the problem \eqref{opt0} whose optimal solution is difficult to obtain, we learn the activity status $\{a_n\}_{n=1}^N$ directly from
the ground-truth training labels based on the received signal model \eqref{received signal}.
We can see from \eqref{received signal} that the received signal $\mathbf{Y}$ is actually
a weighted sum of the active device pilots \textcolor{black}{$\mathbf{S}\mathbf{G}^{\frac{1}{2}}\mathbf{A}=\mathbf{B}\mathbf{A}$}.
To find out which columns of $\mathbf{B}$ contribute to $\mathbf{Y}$,
we need to build a computation mechanism to evaluate the relevance between $\mathbf{Y}$ and $\mathbf{B}$.
Moreover, since each $a_n\in\{0,1\}$ is a discrete variable,
we can view the activity detection as a classification problem, i.e., classifying each $a_n$ \textcolor{black}{as} $0$ or $1$ based on $\mathbf{Y}$ and $\mathbf{B}$.

Specifically, we construct a training data set $\mathcal{D}$ \textcolor{black}{for} supervised \textcolor{black}{learning}, where the $i$-th
training sample is composed of $\left(\mathbf{Y}^{(i)}, \mathbf{B}^{(i)}, \left\{\tilde{a}_n^{(i)}\right\}_{n=1}^N\right)$,
and $\tilde{a}_n^{(i)}$ is the ground-truth label of the $n$-th device's activity status.
The received signal $\mathbf{Y}^{(i)}$ is constructed by substituting a given pair of $\left(\mathbf{B}^{(i)}, \left\{\tilde{a}_n^{(i)}\right\}_{n=1}^N\right)$ into \eqref{received signal}, where
$\mathbf{B}^{(i)}$ is determined by a set of
pilot sequences, large-scale fading gains, and transmit powers,
$ \left\{\tilde{a}_n^{(i)}\right\}_{n=1}^N$ can be generated from Bernoulli distribution,
and the small-scale fading channel $\mathbf{H}$ and the noise $\mathbf{W}$ are sampled from
complex Gaussian distributions. Although $\mathbf{H}$ and $\mathbf{W}$ are used to construct the received signal in the training samples,
they themselves are not explicitly included in the training samples, since $\{a_n\}_{n=1}^N$ should be detected based on $\mathbf{Y}$ and $\mathbf{B}$ without knowing $\mathbf{H}$ and $\mathbf{W}$.

Using the training data set $\mathcal{D}$,
we learn a classifier to infer the
active probability of each device $P_n$
from $\mathbf{Y}$ and $\mathbf{B}$.
Let $f:\mathbb{C}^{L_\text{p}\times M}\times\mathbb{C}^{L_\text{p}\times N}\rightarrow \left[0,1\right]^N$
denote the mapping function from $\left(\mathbf{Y}, \mathbf{B}\right)$ to
$\mathbf{p}\triangleq\left[P_1,\ldots,P_N\right]^T$.
We strive to optimize the mapping function $f\left(\cdot, \cdot\right)$ such that
the difference between the output of the mapping function $\left\{P_n^{(i)}\right\}_{n=1}^N$ and the
ground-truth label $\left\{\tilde{a}_n^{(i)}\right\}_{n=1}^N$ is as close as possible.
For this purpose, we adopt cross entropy \cite{CE} for measuring the discrepancy between
$\left\{P_n^{(i)}\right\}_{n=1}^N$ and $\left\{\tilde{a}_n^{(i)}\right\}_{n=1}^N$,
and learn the classifier by minimizing the following cross entropy based loss function:
\begin{eqnarray}\label{opt1}
\min_{f\left(\cdot, \cdot\right)}
\sum_{i=1}^{\vert\mathcal{D}\vert}\Bigg(\frac{2}{N}\sum_{n=1}^{N}
\Bigg(
\frac{N-K}{N}\tilde{a}_n^{(i)}\log P_n^{(i)}\nonumber\\
+\frac{K}{N}(1-\tilde{a}_n^{(i)})\log \left(1-P_n^{(i)}\right)\Bigg)\Bigg).
\end{eqnarray}
Notice that when the number of active devices is equal to that of inactive devices,
i.e., $K=N-K$, the loss function \eqref{opt1} will reduce to the standard binary cross-entropy loss \cite{CE}, which is widely used for balanced classification in machine learning.
However, due to the sporadic traffics of MTC,
the number of active devices is commonly much less than that of inactive devices,
i..e., $K\ll N-K$, and thus
the standard binary cross-entropy loss will induce overfitting to the inactive class.
In order to avoid overfitting,
we put a much larger weight $\left(N-K\right)/N$ on the loss corresponding to the sporadic active devices while setting a smaller weight $K/N$ on the loss corresponding to the more common inactive devices in~\eqref{opt1}.

\subsection{Parametrization by Neural Network}
\textcolor{black}{To solve problem \eqref{opt1}, we train a neural network (the detailed architecture is given in Section~IV) for parameterizing the mapping function $f\left(\cdot, \cdot\right)$.}
During the training procedure, the neural network learns to adjust its parameters
for minimizing the loss function \eqref{opt1}, so that
the neural network can mimic the optimal mapping function from the received signal $\mathbf{Y}$ and
the scaled pilot matrix $\mathbf{B}$ to the active probability $\mathbf{p}$.
In particular, the neural network parameters can be updated via gradient based methods, e.g., the Adam algorithm \cite{Kingma14}, where the gradients can be automatically computed in any deep learning framework, e.g., Pytorch \cite{Pytorch}. After training, by inputting any $\mathbf{Y}$ and $\mathbf{B}$
into the neural network, we can compute the corresponding output $\mathbf{p}$
via computationally cheap feed-forward operations.
Once $\mathbf{p}$ is
obtained, we can use Bernoulli sampling to obtain the activity status of each device.
Alternatively, we can adopt a threshold $\xi$ to determine the activity status as
$a_n=\mathbb{I}(P_n>\xi)$.

Notice that the training data set is constructed based on the received signal model \eqref{received signal}, where the
ground-truth labels of the device activities are given. This
allows the neural network to mimic the optimal mapping function $f\left(\cdot, \cdot\right)$ directly from the ground-truth labels \cite{Cui2021,Shi2021}. Therefore, there is an opportunity to achieve better detection performance than state-of-the-art covariance based methods that only
obtain a stationary point of the relaxed problems \cite{Jiang2020-2,Chen2021,Ganesan2021}.
Moreover, after the training procedure, the neural networks can infer the solution of any new problem in a real-time manner due to the computationally cheap inference.

\subsection{Limitations of Generic MLPs}
The remaining task is the neural network architecture design for representing the mapping function
$f\left(\cdot, \cdot\right)$.
Although generic MLPs have been widely used for
function representations, they are not tailored to the activity detection problem
mainly due to three reasons.
First, the active probability
of each device $P_n$ should be learned based on the the relevance between
the received signal $\mathbf{Y}$ and the scaled pilot matrix $\mathbf{B}$.
However, generic MLPs
simply concatenate $\mathbf{Y}$ and $\mathbf{B}$ as a single input, and hence
\textcolor{black}{it is difficult to}
extract the relevance between $\mathbf{Y}$ and $\mathbf{B}$.
Second, \textcolor{black}{when any two columns of $\mathbf{B}$
are exchanged and $\mathbf{Y}$ is unchanged, $f\left(\cdot, \cdot\right)$ should output a corresponding permutation of the original
$\mathbf{p}$.} Nevertheless, generic MLPs cannot guarantee the permutation equivariance
for the activity detection problem.
Last but not the least, as the number of devices scales up,
it is highly expected that the neural network is scale adaptable and generalizable to larger numbers of devices than the setting in the training procedure.
Unfortunately, as the input and output dimensions of
generic MLPs are fixed, they are designed
for a pre-defined problem size.
Once the number of devices $N$
has changed, the well-trained MLPs are no longer applicable.

\section{Proposed Heterogeneous Transformer for Representing $f\left(\cdot, \cdot\right)$}
In this section, we propose a customized neural network architecture for representing the mapping function $f\left(\cdot, \cdot\right)$.
Instead of directly applying generic MLPs,
we strive to incorporate properties of the activity detection problem
into the neural network architecture.
In particular, the proposed architecture is capable of extracting the relevance
between the inputs $\mathbf{Y}$ and $\mathbf{B}$, is
permutation equivariant with respect to devices,
and is scale adaptable to different numbers of devices.
Before presenting the proposed architecture, we first briefly review the basic idea of the transformer model.

\subsection{Transformer Model}
Transformer was originally designed for NLP tasks such as machine translation. It adopts an encoder-decoder architecture. The encoder converts the input into a hidden representation by a sequence of encoding layers, while the
decoder produces the output from
the hidden representation by a sequence of decoding layers.
Both the encoder and decoder of transformer are built on the attention mechanism \cite{Vaswani2017},
which extracts the relevance among different input components.
In the context of NLP, the attention mechanism
allows the transformer model to learn from the relevance among different words within a sentence.
Since both the encoder and decoder have a similar structure,
we only review the architecture of the encoder as follows.

The transformer encoder consists of several sequential encoding layers, where
each encoding layer extracts the relevance among the input components.
The generated output of each encoding layer is then passed to the next encoding layer as the input.
Specifically, each encoding layer mainly consists of two blocks:
a multi-head attention (MHA) block that extracts the relevance among different input components, and a component-wise feed-forward (FF) block for additional processing.
Each block further adopts a skip-connection \cite{He16}, which adds an identity mapping to bypass the gradient exploding or vanishing
problem for ease of optimization, and a normalization step \cite{Ba2016}, which
re-scales the hidden representations
to deal with the internal covariate shift in
collective optimization of multiple correlated features.

\begin{figure}[t!]
\begin{center}
  \includegraphics[width=0.5\textwidth]{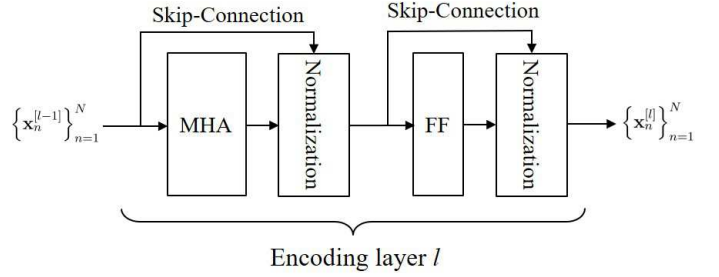}
  \caption{The $l$-th encoding layer of the transformer model.}\label{Transformer}
\end{center}
\vspace{-0.2cm}
\end{figure}

For better understanding, the $l$-th
encoding layer is illustrated in Fig. \ref{Transformer},
where the inputs $\left\{\mathbf{x}_n^{[l-1]}\right\}_{n=1}^{N}$
are passed to the MHA block and the component-wise FF block successively.
The most important module in Fig. \ref{Transformer} is the MHA block, which evaluates the relevance of every pair of components in $\left\{\mathbf{x}_n^{[l-1]}\right\}_{n=1}^{N}$
by scoring how well they match in multiple attention spaces.
By combining all the matching results from different attention spaces,
each layer output is able to capture the relevance among the input components.
In the context of
NLP, this relevance information reflects the importance of each source word and intuitively decides which parts of the source sentence to pay attention to.

\subsection{Architecture of Proposed Heterogeneous Transformer}
The relevance extraction of transformer is appealing for representing the mapping function
$f\left(\cdot, \cdot\right)$, as the device activity should be detected based on the relevance between the received signal $\mathbf{Y}$ and the scaled pilot matrix $\mathbf{B}$.
However, different from the NLP tasks where different words
belong to the same class of features, the inputs $\mathbf{Y}$ and $\mathbf{B}$ in the activity
detection problem have different physical meanings,
and hence should be processed differently.
This observation motivates us to design a heterogeneous
transformer architecture
for representing $f\left(\cdot, \cdot\right)$.

In particular, the proposed heterogenous transformer is composed of an initial embedding layer, $L$ encoding layers, and a decoding layer,
where we use the same set of parameters to process the representations
corresponding to the device pilots, while we process
the representation corresponding to the received signal using another set of
parameters.
The specific architectures of different layers are presented as follows.
For ease of presentation, the scaled pilot matrix is expanded as $\mathbf{B}=\left[\mathbf{b}_1,\ldots,\mathbf{b}_N\right]$,
with the column $\mathbf{b}_n$ corresponding to the $n$-th device.

\noindent\underline{\emph{1) Initial Embedding Layer}}

The initial embedding layer represents the $N$ device pilots $\left\{\mathbf{b}_n\right\}_{n=1}^N$ and the received signal $\mathbf{Y}$ as the input features and then transforms them into an initial representation for the subsequent encoding layers.
The input features are expressed in real-vector forms by separating
the real and imaginary parts. Specifically,
the input features corresponding to
$\left\{\mathbf{b}_n\right\}_{n=1}^N$ are given by
\begin{equation}\label{bin}
\mathbf{x}_n^{\text{in}}=
\left[{\Re\left\{\mathbf{b}_n\right\}}^T,{\Im\left\{\mathbf{b}_n\right\}}^T\right]^T\in\mathbb{R}^{2L_\text{p}},~~\forall n=1,\ldots,N.
\end{equation}
On the other hand, to make the proposed neural network architecture scale adaptable to the number of antennas $M$, we
represent the input features corresponding to $\mathbf{Y}$
by vectorizing the sample covariance matrix $\mathbf{C}\triangleq\mathbf{Y}\mathbf{Y}^H/M$:
\begin{equation}\label{Yin2}
\mathbf{x}_{N+1}^{\text{in}}=
\left[{\Re\left\{\text{vec}\left(\mathbf{C}\right)\right\}}^T,
{\Im\left\{\text{vec}\left(\mathbf{C}\right)\right\}}^T\right]^T\in\mathbb{R}^{2L_\text{p}^2},
\end{equation}
whose dimension is independent of $M$.
In section V, simulation results will be provided to demonstrate the generalizability with respect to different numbers of antennas $M$.

Given the input features $\left\{\mathbf{x}_{n}^{\text{in}}\right\}_{n=1}^{N+1}$, the initial embedding layer applies linear projections to produce the initial embeddings.
Let $d$ denote the dimension of the initial embeddings.
The linear projections are given by
\begin{eqnarray}\label{INI}
\mathbf{x}_{n}^{[0]}=
\begin{cases}
\mathbf{W}^{\text{in}}_{\text{B}}\mathbf{x}_{n}^{\text{in}}+\mathbf{b}^{\text{in}}_{\text{B}},
~~&\forall n=1,\ldots,N,
\cr
\mathbf{W}^{\text{in}}_{\text{Y}}
\mathbf{x}_{N+1}^{\text{in}}+\mathbf{b}^{\text{in}}_{\text{Y}},~~&n=N+1,
\end{cases}
\end{eqnarray}
where $\mathbf{W}^{\text{in}}_{\text{B}}\in\mathbb{R}^{d\times 2L_\text{p}}$ and
$\mathbf{b}^{\text{in}}_{\text{B}}\in\mathbb{R}^{d}$
are the parameters for projecting the input features $\left\{\mathbf{x}_{n}^{\text{in}}\right\}_{n=1}^{N}$,
while $\mathbf{W}^{\text{in}}_{\text{Y}}\in\mathbb{R}^{d\times 2L_\text{p}^2}$
and $\mathbf{b}^{\text{in}}_{\text{Y}}\in\mathbb{R}^{d}$
are the parameters for projecting the input feature $\mathbf{x}_{N+1}^{\text{in}}$.
In \eqref{INI}, the same set of parameters $\left\{\mathbf{W}^{\text{in}}_{\text{B}},
\mathbf{b}^{\text{in}}_{\text{B}}\right\}$
is shared among all devices' pilots, so that
the initial embedding layer is scale adaptable to the number of devices in the actual deployment.  Furthermore, the input feature corresponding to the received signal is processed heterogeneously
by another set of parameters $\left\{\mathbf{W}^{\text{in}}_{\text{Y}},
\mathbf{b}^{\text{in}}_{\text{Y}}\right\}$.
The obtained initial embeddings $\left\{\mathbf{x}_n^{[0]}\right\}_{n=1}^{N+1}$
from \eqref{INI}
are subsequently passed to $L$ encoding layers as follows.

\noindent\underline{\emph{2) Encoding Layers}}

In each encoding layer $l\in\{1,\ldots,L\}$, we adopt the general transformer encoding layer structure in Fig.~\ref{Transformer}. However, the architectures of MHA, FF, and normalization blocks in this work are different from that of the standard transformer, where
all the inputs in a particular
layer are processed using
the same set of parameters.
In contrast, since the device pilots and the received signal
have different physical meanings,
we use one set of parameters to process the inputs $\left\{\mathbf{x}_n^{[l-1]}\right\}_{n=1}^{N}$
(corresponding to the device pilots), and we process
$\mathbf{x}_{N+1}^{[l-1]}$ (corresponding to the received signal) heterogeneously using another set of
parameters.


Specifically, the computation in the $l$-th
encoding layer is described by \eqref{eMHA} and \eqref{eFF},
\newcounter{mytempeqncnt}
\begin{figure*}[!t]
\normalsize
\begin{eqnarray}
\hat{\mathbf{x}}_n^{[l]}&=&
\begin{cases}
\text{BN}_{\text{B}}^l\left(
\mathbf{x}_n^{[l-1]}+\text{MHA}_{\text{B}}^l\left(
\mathbf{x}_n^{[l-1]}, \left\{\mathbf{x}_j^{[l-1]}\right\}_{j=1,j\neq n}^N, \mathbf{x}_{N+1}^{[l-1]}
\right)
\right),& \forall n = 1,\ldots,N,
\cr
\text{BN}_{\text{Y}}^l\left(
\mathbf{x}_{N+1}^{[l-1]}+\text{MHA}_{\text{Y}}^l\left(
\mathbf{x}_{N+1}^{[l-1]}, \left\{\mathbf{x}_j^{[l-1]}\right\}_{j=1}^N
\right)
\right),&n = N+1,
\end{cases}\label{eMHA}\\
\mathbf{x}_n^{[l]}&=&
\begin{cases}
\text{BN}_{\text{B}}^{l}\left(
\hat{\mathbf{x}}_n^{[l]}+\text{FF}_{\text{B}}^{l}\left(
\hat{\mathbf{x}}_n^{[l]}
\right)
\right),~~&\forall n = 1,\ldots,N,
\cr
\text{BN}_{\text{Y}}^{l}\left(
\hat{\mathbf{x}}_{N+1}^{[l]}+\text{FF}_{\text{Y}}^{l}\left(
\hat{\mathbf{x}}_{N+1}^{[l]}
\right)
\right),~~&n = N+1.
\end{cases}\label{eFF}
\end{eqnarray}
\hrulefill
\vspace*{4pt}
\end{figure*}
where $\text{MHA}_{\text{B}}^l$ and $\text{MHA}_{\text{Y}}^l$
denote the MHA computations, $\text{FF}_{\text{B}}^{l}$ and $\text{FF}_{\text{Y}}^{l}$ denote the component-wise FF computations,
$\text{BN}_{\text{B}}^l$ and $\text{BN}_{\text{Y}}^l$ represent the batch normalization (BN) steps \cite{Ioffe15},
and the plus signs represent the skip-connections.
The superscript $l$ indicates that different layers do not share parameters,
while the superscripts $\text{B}$ and $\text{Y}$ mean that
the representations corresponding to the device pilots and the received signal are computed heterogeneously.
In \eqref{eMHA}, $\mathbf{x}_{n}^{[l-1]}$ and $\mathbf{x}_{N+1}^{[l-1]}$ are put \textcolor{black}{outside} of the set $\left\{\mathbf{x}_j^{[l-1]}\right\}_{j=1, j\neq n}^N$,
which implies that each $\mathbf{x}_j^{[l-1]}$ is processed in the same way, while
$\mathbf{x}_{n}^{[l-1]}$ and $\mathbf{x}_{N+1}^{[l-1]}$ are processed in a different way from
$\left\{\mathbf{x}_j^{[l-1]}\right\}_{j=1, j\neq n}^N$.
Next, we explain the computations of
$\text{MHA}_{\text{B}}$, $\text{MHA}_{\text{Y}}$,
$\text{FF}_{\text{B}}$, $\text{FF}_{\text{Y}}$,
$\text{BN}_{\text{B}}$, and $\text{BN}_{\text{Y}}$
in detail.
For notational simplicity, we omit the superscript with respect to $l$ in the following descriptions.

\begin{figure*}[t!]
\begin{center}
 \subfigure[The overall architecture of the MHA block.]{
  \includegraphics[width=0.9\textwidth]{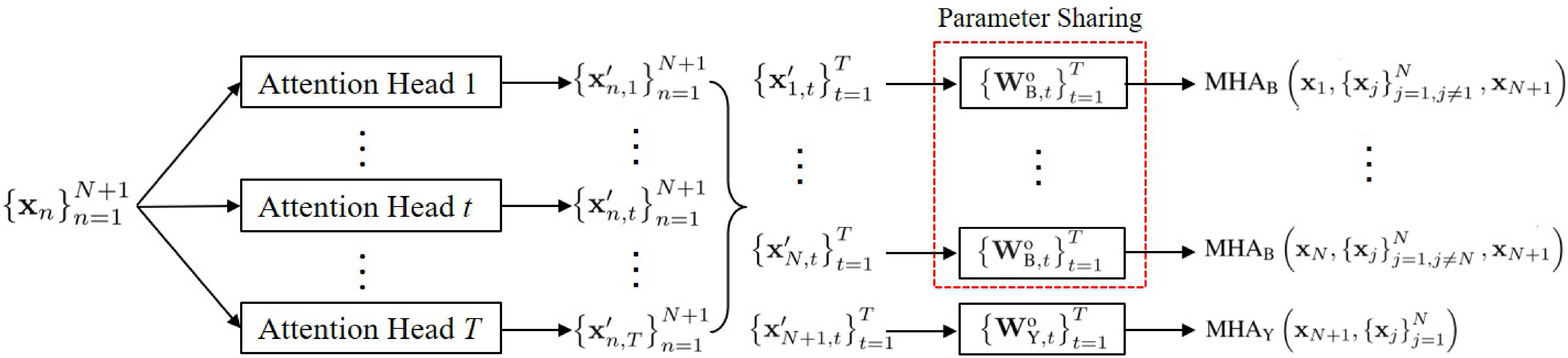}\label{MHA}
  }
  \subfigure[The architecture of attention head $t$ of the MHA block.]{
  \includegraphics[width=0.7\textwidth]{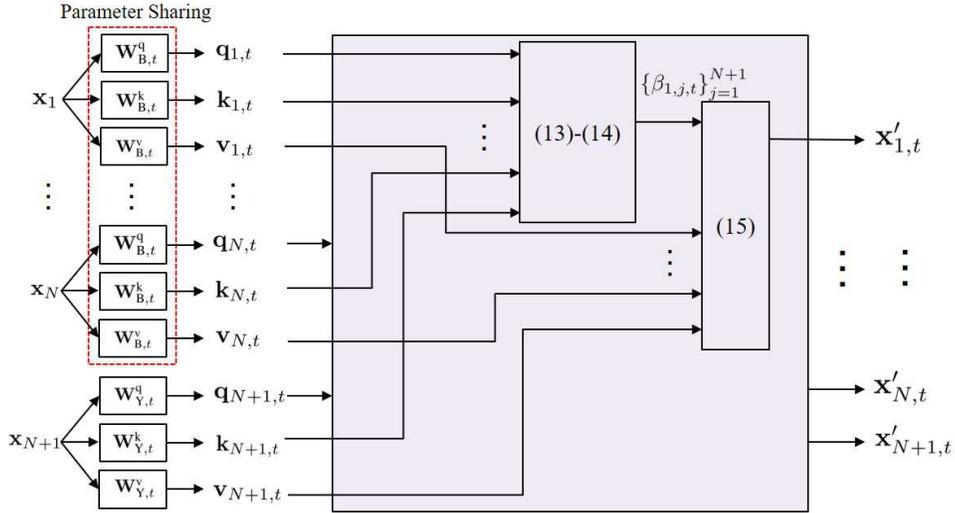}\label{SHA}
}\caption{The architecture of the MHA block.}\label{MHAS}
\end{center}
\vspace{-0.2cm}
\end{figure*}

\emph{a) MHA Computations:} First, we present the MHA computations in \eqref{eMHA},
where we use $T$ attention heads to extract the relevance among the input components (see Fig.~\ref{MHA}).
To describe each attention head, we
define six sets of parameters $\mathbf{W}^{\text{q}}_{\text{B},t}\in\mathbb{R}^{d^{\prime}\times d}$,
$\mathbf{W}^{\text{q}}_{\text{Y},t}\in\mathbb{R}^{d^{\prime}\times d}$,
$\mathbf{W}^{\text{k}}_{\text{B},t}\in\mathbb{R}^{d^{\prime}\times d}$,
$\mathbf{W}^{\text{k}}_{\text{Y},t}\in\mathbb{R}^{d^{\prime}\times d}$,
$\mathbf{W}^{\text{v}}_{\text{B},t}\in\mathbb{R}^{d^{\prime}\times d}$,
and $\mathbf{W}^{\text{v}}_{\text{Y},t}\in\mathbb{R}^{d^{\prime}\times d}$,
where $d^{\prime}$ is the dimension of each attention space and
$t\in\{1,\ldots,T\}$.
For the $t$-th attention head, it computes a query $\mathbf{q}_{n,t}$, a key $\mathbf{k}_{n,t}$, and a value $\mathbf{v}_{n,t}$ for each $\mathbf{x}_n$ (see Fig.~\ref{SHA}):
\begin{eqnarray}\label{eqkv}
\mathbf{q}_{n,t}&=&
\begin{cases}
\mathbf{W}^{\text{q}}_{\text{B},t}\mathbf{x}_n, &\forall n=1,\ldots,N,\label{eq}
\cr
\mathbf{W}^{\text{q}}_{\text{Y},t}\mathbf{x}_{N+1}, &n=N+1,
\end{cases}~~\\
\mathbf{k}_{n,t}&=&
\begin{cases}
\mathbf{W}^{\text{k}}_{\text{B},t}\mathbf{x}_n, &\forall n=1,\ldots,N,\label{ek}
\cr
\mathbf{W}^{\text{k}}_{\text{Y},t}\mathbf{x}_{N+1}, &n=N+1,
\end{cases}~~\\
\mathbf{v}_{n,t}&=&
\begin{cases}
\mathbf{W}^{\text{v}}_{\text{B},t}\mathbf{x}_n, &\forall n=1,\ldots,N,\label{ev}
\cr
\mathbf{W}^{\text{v}}_{\text{Y},t}\mathbf{x}_{N+1}, &n=N+1,
\end{cases}
\end{eqnarray}
where the heterogeneity is reflected in using
parameters $\left\{\mathbf{W}^{\text{q}}_{\text{B},t}, \mathbf{W}^{\text{k}}_{\text{B},t}, \mathbf{W}^{\text{v}}_{\text{B},t}\right\}_{t=1}^T$
to project $\left\{\mathbf{x}_n\right\}_{n=1}^{N}$ (corresponding to the device pilots)
and using parameters $\left\{\mathbf{W}^{\text{q}}_{\text{Y},t}, \mathbf{W}^{\text{k}}_{\text{Y},t}, \mathbf{W}^{\text{v}}_{\text{Y},t}\right\}_{t=1}^T$ to
project $\mathbf{x}_{N+1}$ (corresponding to the received signal)
to different attention spaces.
Then, each attention head computes an attention compatibility $\alpha_{njt}$ for
evaluating how much $\mathbf{x}_n$ is related to $\mathbf{x}_j$:
\begin{eqnarray}\label{compatibility}
\alpha_{n,j,t}=\frac{\mathbf{q}_{n,t}^T\mathbf{k}_{j,t}}{\sqrt{d^{\prime}}},
~~&&\forall n=1,\ldots,N+1,\nonumber\\
~~\forall j=1,\ldots,N+1,
~~&&\forall t=1,\ldots,T,
\end{eqnarray}
and the corresponding attention weight is computed by normalizing $\alpha_{n,j,t}$ in $[0,1]$:
\begin{eqnarray}\label{weight}
\beta_{n,j,t}=\frac{e^{\alpha_{n,j,t}}}{\sum_{j^{\prime}=1}^{N+1}e^{\alpha_{n,j^{\prime},t}}},~~&&\forall n=1,\ldots,N+1,\nonumber\\
~~\forall j=1,\ldots,N+1,
~~&&\forall t=1,\ldots,T.
\end{eqnarray}
With the attention weight $\beta_{n,j,t}$ scoring the relevance between $\mathbf{x}_n$ and $\mathbf{x}_j$, the attention value of $\mathbf{x}_n$ at the $t$-th attention head
is computed as a weighted sum\footnote{Notice that
the summation of \eqref{attention} is taken over $j$ rather than $n$.
Therefore, $\mathbf{x}^{\prime}_{n,t}$ serves as the attention value at the $t$-th attention head corresponding to $\mathbf{x}_n$.}:
\begin{equation}\label{attention}
\mathbf{x}^{\prime}_{n,t}=\sum_{j=1}^{N+1} \beta_{n,j,t}\mathbf{v}_{j,t},~~\forall n=1,\ldots,N+1.
\end{equation}
Finally, by combining the attention values from $T$ attention heads
with parameters $\left\{\mathbf{W}^{\text{o}}_{\text{B},t}\in\mathbb{R}^{d\times d^{\prime}}\right\}_{t=1}^T$
and $\left\{\mathbf{W}^{\text{o}}_{\text{Y},t}\in\mathbb{R}^{d\times d^{\prime}}\right\}_{t=1}^T$,
$\left\{\mathbf{x}^{\prime}_{n,t}\right\}_{n=1}^{N+1}$
are projected back to $d$-dimensional vectors and
we obtain the MHA computation results (see Fig.~\ref{MHA}):
\begin{eqnarray}
\text{MHA}_{\text{B}}\left(
\mathbf{x}_n, \left\{\mathbf{x}_j\right\}_{j=1,j\neq n}^N, \mathbf{x}_{N+1}
\right)=\sum_{t=1}^T \mathbf{W}^{\text{o}}_{\text{B},t}\mathbf{x}^{\prime}_{n,t},~~~&&\nonumber\\~~\forall n=1,\ldots,N,&&\label{eMHAvalue1}\\
\text{MHA}_{\text{Y}}\left(
\mathbf{x}_{N+1}, \left\{\mathbf{x}_j\right\}_{j=1}^N
\right)=\sum_{t=1}^T \mathbf{W}^{\text{o}}_{\text{Y},t}\mathbf{x}^{\prime}_{{N+1},t},&&\label{eMHAvalue2}
\end{eqnarray}
where \eqref{eMHAvalue1} and \eqref{eMHAvalue2}
correspond to the
projections for the device pilots
and the received signal, respectively.
Notice that $\mathbf{x}_{n}$ and
$\mathbf{x}_{N+1}$ are put \textcolor{black}{outside} of $\left\{\mathbf{x}_j\right\}_{j=1,j\neq n}^{N}$ in \eqref{eMHAvalue1},
because $\mathbf{x}^{\prime}_{n,t}$ is computed by
processing
each $\mathbf{x}_j$ using $\left\{\mathbf{W}^{\text{k}}_{\text{B},t},
\mathbf{W}^{\text{v}}_{\text{B},t}\right\}$ in the same way, while by processing
$\mathbf{x}_{n}$ and
$\mathbf{x}_{N+1}$ using $\left\{\mathbf{W}^{\text{q}}_{\text{B},t}, \mathbf{W}^{\text{k}}_{\text{B},t}, \mathbf{W}^{\text{v}}_{\text{B},t}\right\}$ and
$\left\{\mathbf{W}^{\text{k}}_{\text{Y},t},
\mathbf{W}^{\text{v}}_{\text{Y},t}\right\}$, respectively.

\emph{b) FF Computations:} Next, we present the computations of $\text{FF}_{\text{B}}$ and $\text{FF}_{\text{Y}}$ in \eqref{eFF}, which
adopt a two-layer MLP with a $d_{\text{f}}$-dimensional hidden layer using the ReLu
activation:
\begin{eqnarray}
\text{FF}_{\text{B}}\left(
\hat{\mathbf{x}}_n
\right)=
\mathbf{W}^{\text{f}}_{\text{B},2}\text{ReLu}\left(\mathbf{W}^{\text{f}}_{\text{B},1}\hat{\mathbf{x}}_n+\mathbf{b}^{\text{f}}_{\text{B},1}
\right)+\mathbf{b}^{\text{f}}_{\text{B},2},~~~\nonumber\\\forall n=1,\ldots,N,&&\label{eFFvalue1}\\
\text{FF}_{\text{Y}}\left(
\hat{\mathbf{x}}_{N+1}
\right)=
\mathbf{W}^{\text{f}}_{\text{Y},2}\text{ReLu}\left(\mathbf{W}^{\text{f}}_{\text{Y},1}\hat{\mathbf{x}}_{N+1}+\mathbf{b}^{\text{f}}_{\text{Y},1}
\right)+\mathbf{b}^{\text{f}}_{\text{Y},2},&&\label{eFFvalue2}
\end{eqnarray}
where $\left\{\hat{\mathbf{x}}_n\right\}_{n=1}^{N+1}$ is the output of \eqref{eMHA}, and
$\mathbf{W}^{\text{f}}_{\text{B},1}\in\mathbb{R}^{d_{\text{f}}\times d}$,
$\mathbf{b}^{\text{f}}_{\text{B},1}\in\mathbb{R}^{d_{\text{f}}}$,
$\mathbf{W}^{\text{f}}_{\text{B},2}\in\mathbb{R}^{d\times d_{\text{f}}}$,
$\mathbf{b}^{\text{f}}_{\text{B},2}\in\mathbb{R}^{d}$,
$\mathbf{W}^{\text{f}}_{\text{Y},1}\in\mathbb{R}^{d_{\text{f}}\times d}$,
$\mathbf{b}^{\text{f}}_{\text{Y},1}\in\mathbb{R}^{d_{\text{f}}}$,
$\mathbf{W}^{\text{f}}_{\text{Y},2}\in\mathbb{R}^{d\times d_{\text{f}}}$,
and $\mathbf{b}^{\text{f}}_{\text{Y},2}\in\mathbb{R}^{d}$
are the parameters to be optimized during the training procedure.
In \eqref{eFFvalue1}-\eqref{eFFvalue2}, the heterogeneity is maintained since
we use the same set of parameters $\left\{\mathbf{W}^{\text{f}}_{\text{B},1},
\mathbf{b}^{\text{f}}_{\text{B},1},
\mathbf{W}^{\text{f}}_{\text{B},2},
\mathbf{b}^{\text{f}}_{\text{B},2}\right\}$
to process $\left\{\hat{\mathbf{x}}_n\right\}_{n=1}^{N}$
(corresponding to the device pilots), while we process $\hat{\mathbf{x}}_{N+1}$
(corresponding to the received signal)
by another set of parameters $\left\{\mathbf{W}^{\text{f}}_{\text{Y},1},
\mathbf{b}^{\text{f}}_{\text{Y},1},
\mathbf{W}^{\text{f}}_{\text{Y},2},
\mathbf{b}^{\text{f}}_{\text{Y},2}\right\}$.

\emph{c) BN Computations:} For the BN computations in \eqref{eMHA} and \eqref{eFF},
it computes the statistics over a batch of training samples.
Specifically, let
$\left\{\tilde{\mathbf{x}}_n^{(i)}\in\mathbb{R}^{d}\right\}_{i=1}^{I_{\text{b}}}$
denote a mini-batch of training samples for BN computation.
The BN statistics are calculated as
\begin{eqnarray}\label{BNstatistics}
\boldsymbol{\nu}_n &=& \frac{1}{I_{\text{b}}}\sum_{i=1}^{I_{\text{b}}}\tilde{\mathbf{x}}_n^{(i)},~~\forall n=1,\ldots,N+1,\\
\boldsymbol{\Gamma}_n &=& {\left(\frac{1}{I_{\text{b}}}\sum_{i=1}^{I_{\text{b}}}
{\boldsymbol{\Lambda}_n^{(i)}}\right)}^{\frac{1}{2}},~~\forall n=1,\ldots,N+1,
\end{eqnarray}
where ${\boldsymbol{\Lambda}_n^{(i)}}$ is a diagonal matrix with the diagonal
being $\left(\tilde{\mathbf{x}}_n^{(i)}-\boldsymbol{\nu}_n\right)\odot\left(\tilde{\mathbf{x}}_n^{(i)}-\boldsymbol{\nu}_n\right)$.
Then, the normalization results corresponding to
the device pilots and the received signal are respectively given by
\begin{eqnarray}
\text{BN}_{\text{B}}\left(
\tilde{\mathbf{x}}_n^{(i)}
\right)=\mathbf{w}^{\text{bn}}_{\text{B}}\odot\left(\boldsymbol{\Gamma}_n^{-1}\left(\tilde{\mathbf{x}}_n^{(i)}-\boldsymbol{\nu}_n\right)\right)
+\mathbf{b}^{\text{bn}}_{\text{B}},~~~~~~~~&&
\nonumber\\~~\forall n=1,\ldots,N,&&\label{BN1}\\
\text{BN}_{\text{Y}}\left(
\tilde{\mathbf{x}}_{N+1}^{(i)}
\right)=\mathbf{w}^{\text{bn}}_{\text{Y}}\odot\left(\boldsymbol{\Gamma}_{N+1}^{-1}\left(\tilde{\mathbf{x}}_{N+1}^{(i)}-\boldsymbol{\nu}_{N+1}\right)\right)
+\mathbf{b}^{\text{bn}}_{\text{Y}},&&\label{BN2}
\end{eqnarray}
where
$\mathbf{w}^{\text{bn}}_{\text{B}}\in\mathbb{R}^{d}$,
$\mathbf{b}^{\text{bn}}_{\text{B}}\in\mathbb{R}^{d}$,
$\mathbf{w}^{\text{bn}}_{\text{Y}}\in\mathbb{R}^{d}$,
and $\mathbf{b}^{\text{bn}}_{\text{Y}}\in\mathbb{R}^{d}$
are the parameters to be optimized during the training procedure.

\noindent\underline{\emph{3) Decoding Layer}}

After the $L$ encoding layers,
the produced hidden representations $\left\{\mathbf{x}_n^{[L]}\right\}_{n=1}^{N+1}$
are further passed to a decoding layer to output the final mapping result.
The proposed decoding layer consists of a contextual block and an output block.
The contextual block applies an MHA block to
compute a context vector $\mathbf{x}^{\text{c}}$, which
is a weighted sum of the components in $\left\{\mathbf{x}_n^{[L]}\right\}_{n=1}^{N+1}$:
\begin{equation}\label{context}
\mathbf{x}^{\text{c}}=\text{MHA}_{\text{C}}\left(
\mathbf{x}_{N+1}^{[L]}, \left\{\mathbf{x}_n^{[L]}\right\}_{n=1}^N
\right),
\end{equation}
where $\text{MHA}_{\text{C}}$ is similar to $\text{MHA}_{\text{Y}}$ in \eqref{eMHAvalue2}, but using different parameters
$\mathbf{W}^{\text{q}, \text{c}}_{t}\in\mathbb{R}^{d^{\prime}\times d}$,
$\mathbf{W}^{\text{k}, \text{c}}_{\text{B},t}\in\mathbb{R}^{d^{\prime}\times d}$,
$\mathbf{W}^{\text{k}, \text{c}}_{\text{Y},t}\in\mathbb{R}^{d^{\prime}\times d}$,
$\mathbf{W}^{\text{v}, \text{c}}_{\text{B},t}\in\mathbb{R}^{d^{\prime}\times d}$,
$\mathbf{W}^{\text{v}, \text{c}}_{\text{Y},t}\in\mathbb{R}^{d^{\prime}\times d}$,
$\mathbf{W}^{\text{o}, \text{c}}_{t}\in\mathbb{R}^{d\times d^{\prime}}$, $t\in\{1,\ldots,T\}$.
In particular,
$\left\{\mathbf{W}^{\text{k}, \text{c}}_{\text{B},t},
\mathbf{W}^{\text{v}, \text{c}}_{\text{B},t}\right\}_{t=1}^T$ is used to
process $\left\{\mathbf{x}_n^{[L]}\right\}_{n=1}^N$
(corresponding to the device pilots),
and $\left\{\mathbf{W}^{\text{q}, \text{c}}_{t},
\mathbf{W}^{\text{k}, \text{c}}_{\text{Y},t}, \mathbf{W}^{\text{v}, \text{c}}_{\text{Y},t},
\mathbf{W}^{\text{o}, \text{c}}_{t}\right\}_{t=1}^T$ is used to process $\mathbf{x}_{N+1}^{[L]}$
(corresponding to the received signal).
The specific expression of $\text{MHA}_{\text{C}}$ is shown in Appendix \ref{MHAC}.
Each weight in $\mathbf{x}^{\text{c}}$ reflects the importance of each component in
$\left\{\mathbf{x}_n^{[L]}\right\}_{n=1}^{N+1}$. Therefore, the context vector $\mathbf{x}^{\text{c}}$
intuitively decides which device pilots to pay attention to based
on the received signal.


With $\mathbf{x}^{\text{c}}$,
the output block decides the final
output, i.e., the active probability of each device,
by scoring how well the context vector $\mathbf{x}^{\text{c}}$ and each
$\mathbf{x}_n^{[L]}$, $n\in\{1,\ldots,N\}$ match.
The relevance between the context vector $\mathbf{x}^{\text{c}}$
and each $\mathbf{x}_n^{[L]}$ is evaluated
by
\begin{equation}\label{outcompatibility}
\alpha_{n}^{\text{out}}=C\tanh\left(
\frac{\left(\mathbf{x}^{\text{c}}\right)^T\mathbf{W}_{\text{out}}\mathbf{x}_n^{[L]}}{\sqrt{d}}\right),~~\forall n=1,\ldots,N,
\end{equation}
where $\mathbf{W}_{\text{out}}\in\mathbb{R}^{d\times d}$ is a parameter to be optimized during the training procedure,
and $C$ is a tuning hyperparameter that controls $\alpha_{n}^{\text{out}}$ in a reasonable range.
Finally, the active probability of each device is computed
by normalizing $\alpha_{n}^{\text{out}}$ in $[0,1]$:
\begin{equation}\label{out}
P_n=
\text{OUT}\left(\mathbf{x}^{\text{c}}, \mathbf{x}_n^{[L]}
\right)=
\frac{1}{1+e^{-\alpha_{n}^{\text{out}}}},~~\forall n=1,\ldots,N.
\end{equation}

\subsection{Key Properties and Insights}
The proposed heterogenous transformer for representing
$f\left(\cdot, \cdot\right)$ has been specified as an initial embedding layer, $L$ encoding layer, and a decoding layer as shown in \eqref{bin}-\eqref{out}.
We examine some key properties of the proposed architecture for the activity detection problem as follows.

\begin{enumerate}[a)]
\item
\emph{Relevance Extraction Between Device Pilots and Received Signal:}
Both the proposed encoding and decoding layers are built on MHA as shown in \eqref{eMHA} and \eqref{context}, respectively.
The MHA computation is naturally a weighted sum as shown in \eqref{attention}. The attention weight $\beta_{n,j,t}$
is the normalization of the attention compatibility $\alpha_{n,j,t}$ in \eqref{compatibility}, which scores how well each pair of $\mathbf{x}_n$ and $\mathbf{x}_j$ match.
Therefore, with the attention weight $\beta_{n,j,t}$ reflecting the importance of each $\mathbf{x}_j$ with respect to $\mathbf{x}_n$, each encoding layer learns the relevance
among different device pilots and the received signal.
In the decoding layer, the captured relevance is further used
to compute the context vector $\mathbf{x}^{\text{c}}$ in \eqref{context},
which finally extracts the relevance between each device pilot and the received signal,
and decides which device pilots to pay attention to based on the extracted relevance.

\item
\emph{Permutation Equivariant with Respect to Devices:}
As shown in \eqref{INI},
the initial embeddings of the devices pilots $\left\{\mathbf{x}_n^{[0]}\right\}_{n=1}^{N}$
are computed using the same parameters $\mathbf{W}^{\text{in}}_{\text{B}}$ and
$\mathbf{b}^{\text{in}}_{\text{B}}$. Therefore,
if any two device pilots $\mathbf{b}_i$ and $\mathbf{b}_j$ are exchanged,
the initial embeddings $\mathbf{x}_i^{[0]}$ and $\mathbf{x}_j^{[0]}$ will be automatically exchanged as well.
Similarly, since the encoding layers produce $\left\{\mathbf{x}_n^{[L]}\right\}_{n=1}^{N}$
through the same computations $\text{MHA}_{\text{B}}^l$, $\text{FF}_{\text{B}}^l$, and $\text{BN}_{\text{B}}^l$,
and the decoding layer
produces $\left\{P_n\right\}_{n=1}^{N}$ by the same parameter
$\mathbf{W}_{\text{out}}$, we can conclude that the final output $P_i$ and $P_j$ will also be exchanged.
This implies that
the proposed architecture is permutation equivariant with respect to devices.

\item
\emph{Scale Adaptable and Generalizable to Different Numbers of Devices:}
In all the layers of the proposed heterogeneous transformer, the
representations of different
device pilots are produced with the same architecture
using the same set of parameters.
Therefore, the dimensions of parameters that
require to be optimized during the training procedure
are independent of the number of devices.
This \textcolor{black}{scale adaptability} empowers the whole architecture
to be readily applied to different numbers of devices,
and hence generalizable to different numbers of devices.
\end{enumerate}


\subsection{Learning Procedure}
So far, we have presented the architecture and key properties of the proposed heterogeneous transformer.
Next, we \textcolor{black}{show the} learning procedure to optimize the parameters of heterogeneous transformer
for device activity detection
in Algorithm~1, which consists of a training procedure and a test procedure.
\textcolor{black}{As shown in lines 8-10,
we adopt a learning rate decay strategy to
\textcolor{black}{accelerate the training procedure} \cite{You2019}.
In particular, the learning rate $\eta$ is decreased by a factor of $\beta$ after $N_{\text{d}}$ training epochs.
During the test procedure, we adopt two metrics to assess the performance of
device activity detection,
i.e., the probability of missed detection (PM) and the probability of false alarm (PF) \cite{LiuLiang2018,LiuLiang2018-1,LiuLiang2018-2,Chen2018},
which are respectively given by
\begin{eqnarray}
\text{PM}&=&1-\frac{\sum_{n=1}^Na_n\tilde{a}_n}{\sum_{n=1}^N\tilde{a}_n},\label{PM}\\
\text{PF}&=&\frac{\sum_{n=1}^Na_n\left(1-\tilde{a}_n\right)}{\sum_{n=1}^N(1-\tilde{a}_n)}.\label{PF}
\end{eqnarray}
In \eqref{PM} and \eqref{PF},
$\tilde{a}_n$ is the ground-truth device activity, and the detected
activity status
$a_n=\mathbb{I}(P_n>\xi)$, where
$\xi$ is a threshold that
continually increases in $[0, 1]$ to realize a trade-off between PM and PF.}

\begin{algorithm}[t!]
\caption{Learning Procedure for Activity Detection}
\begin{algorithmic}[1]\footnotesize
\State \underline{\textbf{Training Procedure:}}\\
\textbf{Input:} number of epochs $N_{\text{e}}$,
steps per epoch $N_{\text{s}}$,
batch size $N_{\text{b}}$,
and learning rate decay epoch  $N_{\text{d}}$ and
factor $\beta$\\
\textbf{Initialize:} learning rate $\eta$\\
$\textbf{for}$ $\text{epoch} = 1, \ldots, N_{\text{e}}$\\
~~~~\textbf{for} $\text{step} = 1, \ldots, N_{\text{s}}$\\
\begin{enumerate}[]
\item
~~~~a) Generate a batch of $N_{\text{b}}$ samples
\item
~~~~b) Compute the mini-batch gradient of the loss function \eqref{opt1} over\\
~~~~~~~the parameters of heterogeneous transformer
\item
~~~~c) Update the parameters by a gradient descent step using the \\
~~~~~~~Adam optimizer with learning rate $\eta$
\end{enumerate}\\
~~~~\textbf{end}\\
~~~~\textbf{if} $\text{epoch} == N_{\text{d}}$\\
~~~~~~~~$\eta\leftarrow\beta\eta$\\
~~~~\textbf{end}\\
$\textbf{end}$\\
\textbf{Output:} heterogeneous transformer with optimized parameters\\
\underline{\textbf{Test Procedure:}}\\
\textbf{Input:}  $N_{\text{t}}$ test samples\\
Compute the output of the trained heterogeneous transformer $P_n^{(i)}$,
$n=1,\ldots,N$, $i=1,\ldots,N_{\text{t}}$\\
Determine the activity status of each device as $a_n^{(i)}=\mathbb{I}(P_n^{(i)}>\xi)$, $n=1,\ldots,N$, $i=1,\ldots,N_{\text{t}}$\\
\textbf{Output:} (PM, PF) pairs under different $\xi$
\end{algorithmic}
\end{algorithm}

\section{Simulation Results}
In this section, simulation results are provided for demonstrating
the benefits of the proposed learning based method, which
adopts the problem formulation for learning in Section~III and
the heterogeneous transformer architecture in Section~IV.


\subsection{Simulation Setting}
We consider an uplink MIMO system with IoT devices uniformly distributed within
a cell with a $250$-meter radius, and the ratio of the active devices
to the total devices is $0.1$.
Both the training and test samples are generated as follows. The pilot sequence of each device is
an independently generated
complex Gaussian distributed vector with i.i.d. elements and each element is
with zero mean and unit variance.
The path-loss of the channel
is $128.1+37.6\log_{10}{D_n}$ in dB, where $D_n$ is the distance in kilometers between the $n$-th device
and the BS.
In order to reduce the channel gain variations among
different devices especially for the cell-edge devices,
the transmit power of each device is controlled as $p_k=p_{\max}\frac{g_{\min}}{g_k}$ \cite{Senel2018},
where $p_{\max}$ is the maximum transmit power and
$g_{\min}$ is the minimum large-scale \textcolor{black}{channel gain} in the cell.
The background Gaussian noise power at the BS is
$-169$ dBm/Hz over a $10$ MHz bandwidth.
\textcolor{black}{The received signal is generated
according to \eqref{received signal}, where the activity of each device is
generated from Bernoulli distribution and
used as the ground-truth label for the training samples.}

For the proposed heterogeneous transformer architecture,
the number of encoding layers is set as $L=5$,
the encoding size of each encoding layer is $d=128$,
the number of attention heads in the MHA block is $T=8$,
the dimension of each attention space is $d^{\prime}=32$,
and the size of hidden layer of the component-wise FF block is $d_{\text{f}}=512$.
In the decoding layer, the tuning hyperparameter that controls the result of \eqref{outcompatibility}
is set as $C=10$.

\textcolor{black}{
During the training procedure,
we apply the Adam optimizer in the framework of Pytorch \cite{Pytorch}
to update the parameters of the proposed heterogeneous transformer. In Algorithm~1,
the number of training epochs is set as $N_{\text{e}}=100$.
The number of gradient descent steps in each epoch is
$N_{\text{s}}=5000$ and each step is updated using a batch of $N_{\text{b}}=256$ training samples.
Consequently, the total number of training samples is $1.28\times10^{6}$.
The learning rate is initialized as $\eta=10^{-4}$.
For the learning rate decay strategy,
we decrease $\eta$ by a factor of $\beta=0.1$ after $N_{\text{d}}=90$ and $97$ training epochs, respectively.
After the training procedure, the activity detection performance
is evaluated over $N_{\text{t}}=5000$ test samples \textcolor{black}{in terms of PM and PF
as shown in \eqref{PM} and \eqref{PF}.}
}

\subsection{Performance Evaluation}
\begin{figure}[t!]
\begin{center}
  \includegraphics[width=0.48\textwidth]{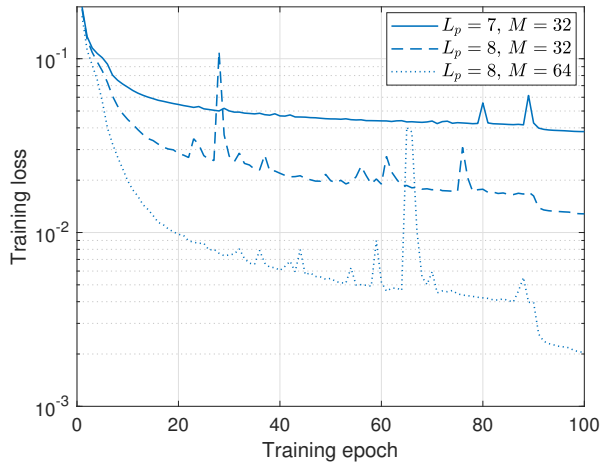}
  \caption{The training loss of Algorithm 1 for updating the parameters of heterogeneous transformer.}\label{con}
\end{center}
\vspace{-0.2cm}
\end{figure}

First, we show the training loss of Algorithm 1 for updating the parameters of heterogeneous transformer.
During the training procedure, the number of devices is set as $N=100$ and
the maximum transmit power is $p_{\max}=23$ dBm.
The length of each pilot sequence is set as $L_{\text{p}}=7$ or $8$, and
the number of BS-antennas is set as $M=32$ or $64$, respectively.
The training losses versus epochs under different settings are illustrated in Fig. \ref{con}.
It can be seen that the training losses generally decrease
as the training epoch increases.
In particular, due to the learning rate decay, the training losses have a sudden decrease
in the $90$-th training epoch, which
demonstrates the effectiveness of learning rate decay in
speeding up the training procedure. We also observe from
Fig. \ref{con} that
the training performance can be improved by increasing the
the length of pilot sequence or equipping with a larger number of antennas at the BS.

\begin{figure}[t!]
\begin{center}
  \includegraphics[width=0.48\textwidth]{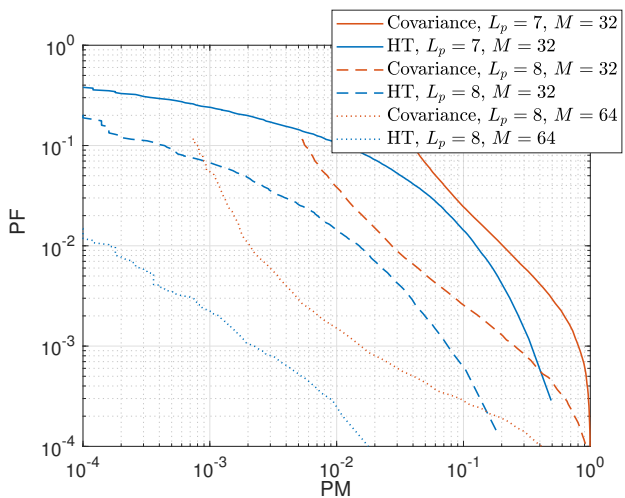}
  \caption{The test performance comparison with state-of-the-art covariance approach in terms of PM and PF.}\label{com}
\end{center}
\vspace{-0.2cm}
\end{figure}

Then, we test the corresponding activity detection performance of the well-trained
heterogeneous transformer in terms of PM and PF.
For comparison, we also provide the simulation results of state-of-the-art covariance approach\textcolor{black}{, which solves problem \eqref{opt0} with the coordinate descent method \cite{Chen2021,Ganesan2021}.}
The PM-PF trade-offs are shown in Fig. \ref{com},
where the proposed learning based method using heterogeneous transformer is termed as HT,
while state-of-the-art covariance approach is termed as Covariance.
It can be seen that the proposed \textcolor{black}{method} always achieves better
PM-PF trade-offs than that of \textcolor{black}{the covariance approach} under different settings.
In particular, when $L_{\text{p}}=8$
and $M=64$, the PM of \textcolor{black}{the proposed method}
is about $10$ times lower than that of \textcolor{black}{the covariance approach} under the same PF.
\textcolor{black}{This is because the proposed method utilizes
neural network to mimic the optimal mapping function directly from the ground-truth
training labels, which provide the opportunity to achieve better detection performance than the covariance approach that only finds a stationary point of a relaxed problem of \eqref{opt0}.}

\textcolor{black}{
To further demonstrate the benefits of
the properties incorporated in the proposed heterogeneous transformer,
we also perform the same tasks using MLPs for comparison. In particular, we train different
MLPs with $4$-$10$ hidden layers and $256$-$1024$ hidden nodes using the ReLu activation.
However, without a custom design, none of the MLPs can provide a better detection performance than even a random guess (and hence are omitted in Fig.~\ref{com}). This verifies the performance gains of the proposed heterogeneous transformer architecture for the activity detection problem.}

The average computation times of the approaches over the test samples are compared in Table~I,
where the \textcolor{black}{covariance approach is termed as Covariance, and the proposed method is termed as
either HT CPU or HT GPU, depending on whether CPU or GPU is used.
In particular, both the covariance approach}
and HT CPU are run on Intel(R) Xeon(R) CPU @ 2.20GHz, while
HT GPU is run on Tesla T4. We can see that
the average computation time of HT CPU is
about $100$ times shorter than that of \textcolor{black}{the covariance approach}.
Moreover, HT GPU achieves a remarkable running speed, with a running time
over $10^{5}$ times shorter than that of the covariance approach.
This demonstrates the superiority of the proposed learning based method for real-time implementation
compared with the covariance approach that is based on iterative computations.

\begin{table}[t!]\caption{Average computation time comparison among different approaches}
\begin{center}\footnotesize
\begin{tabular}{ccccc}
\hline
     & Covariance &  HT CPU & HT GPU \\ \hline
 $L_{\text{p}}=7$, $M=32$     & $6.34\times10^{-1}$ s  &  $6.31\times10^{-3}$ s & $1.60\times10^{-6}$ s \\ \hline
 $L_{\text{p}}=8$, $M=32$  & $6.24\times10^{-1}$ s  &   $6.32\times10^{-3}$ s & $1.63\times 10^{-6}$ s\\  \hline
 $L_{\text{p}}=8$, $M=64$  & $6.33\times10^{-1}$ s &  $6.55\times10^{-3}$ s & $2.13\times 10^{-6}$ s\\ \hline
\end{tabular}
\end{center}
\vspace{-0.2cm}
\end{table}


\subsection{Generalizability}
\begin{figure}[t!]
\begin{center}
 \subfigure[PM versus number of devices.]{
  \includegraphics[width=0.48\textwidth]{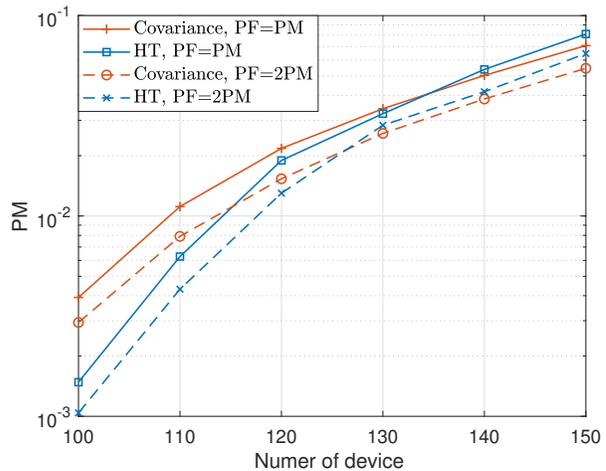}\label{PM_N}
  }
  \subfigure[Average computation time versus number of devices.]{
  \includegraphics[width=0.48\textwidth]{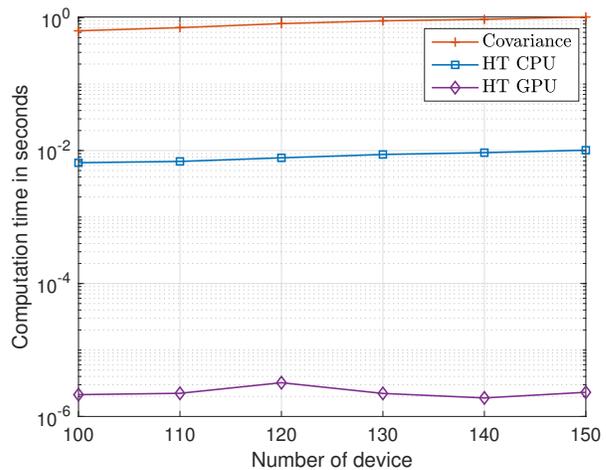}\label{T_N}
}\caption{Generalization to different numbers of devices.}\label{M}
\end{center}
\vspace{-0.2cm}
\end{figure}

Next, we demonstrate the generalizability of the proposed method
with respect to different numbers of devices, BS-antennas, and different SNRs.
Unless otherwise specified, the length of pilot sequence is set as $L_{\text{p}}=8$
in the following simulations.
We begin by training a heterogeneous transformer, where the device number of the training samples is fixed as $N=100$.
However, we test the activity detection performance under different device numbers from $100$ to $150$.
The number of BS-antennas is set as $M=64$ and the maximum transmit power is $p_{\max}=23$ dBm.
Due to the trade-off between PM and PF, we provide the PM
when PF = PM and PF = $2$PM respectively, by appropriately setting the threshold $\xi$.
\textcolor{black}{In the following figures, ``Covariance, PF = PM'' and ``Covariance, PF = $2$PM''
denote the PM of the covariance approach when  PF = PM and PF = $2$PM respectively,
while ``HT, PF = PM'' and ``HT, PF = $2$PM'' represent the PM of the proposed method
when PF = PM and PF = $2$PM respectively.
The activity detection performance and average computation time versus number of devices are illustrated in Fig.~\ref{PM_N} and Fig.~\ref{T_N}, respectively.}
We can see from Fig.~\ref{PM_N} that while the activity detection performances of different approaches become worse
as the number of devices $N$ increases, the PM of the proposed method is still
comparable with that of the covariance approach when $N$ is increased from $100$ to $150$.
This demonstrates that the proposed method generalizes well to different numbers of devices.
On the other hand, Fig.~\ref{T_N} shows that as $N$ increases,
the average computation times of both the covariance approach and the proposed method on CPU
are linearly increased.
However, due to the the parallel computation of GPU, the average computation time of the proposed method on GPU is nearly a constant, i.e., about $2\times10^{-6}$ second,
which is much shorter than that of other approaches.

\begin{figure}[t!]
\begin{center}
 \subfigure[PM versus number of BS-antennas.]{
  \includegraphics[width=0.48\textwidth]{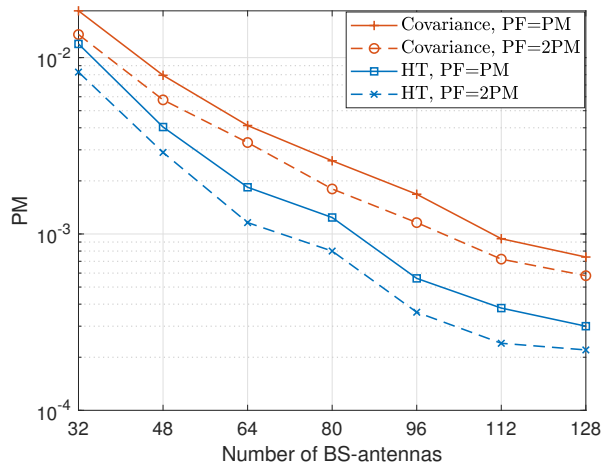}\label{PM_M}
  }
  \subfigure[Average computation time versus number of BS-antennas.]{
  \includegraphics[width=0.48\textwidth]{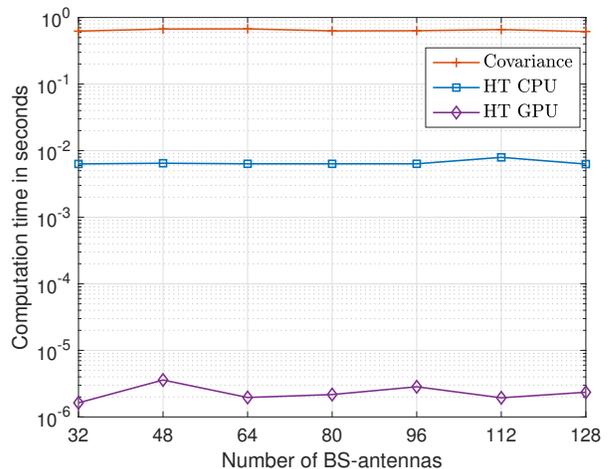}\label{T_M}
}\caption{Generalization to different numbers of BS-antennas.}\label{M}
\end{center}
\vspace{-0.2cm}
\end{figure}

We further demonstrate the generalizability of the proposed method with respect to different numbers of BS-antennas.
To this end, we train a heterogeneous transformer by fixing the
number of BS-antennas as $M=32$, and then test its activity detection performance
under different numbers of BS-antennas
from $32$ to $128$. The number of devices is $N=100$ and the maximum transmit power is $p_{\max}=23$ dBm.
The performance comparisons in terms of
PM and average computation time are illustrated
in Fig.~\ref{PM_M} and Fig.~\ref{T_M}, respectively.
Figure~\ref{PM_M} shows that as the number of BS-antennas increases,
the proposed method always achieves much lower PM than that of the covariance approach.
\textcolor{black}{Although the heterogeneous transformer
is trained under $M=32$, when we test the detection performance
under $M=128$, the PM of the proposed method}
is \textcolor{black}{still} $2$ times lower than that of the covariance approach for both PF = PM and PF = $2$PM.
This demonstrates that the proposed method generalizes well to larger numbers of BS-antennas.
On the other hand,
Fig. \ref{T_M} shows that the average computation time of the proposed method on CPU is
about $100$ times shorter than that of the covariance approach,
and the proposed method on GPU even achieves a $10^{5}$ times faster running speed than that of the covariance approach.

\begin{figure}[t!]
\begin{center}
 \subfigure[PM versus maximum transmit power.]{
  \includegraphics[width=0.48\textwidth]{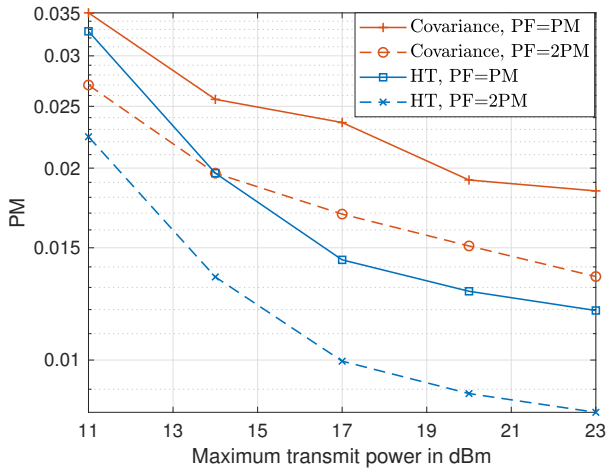}\label{PM_P}
  }
  \subfigure[Average computation time versus maximum transmit power.]{
  \includegraphics[width=0.48\textwidth]{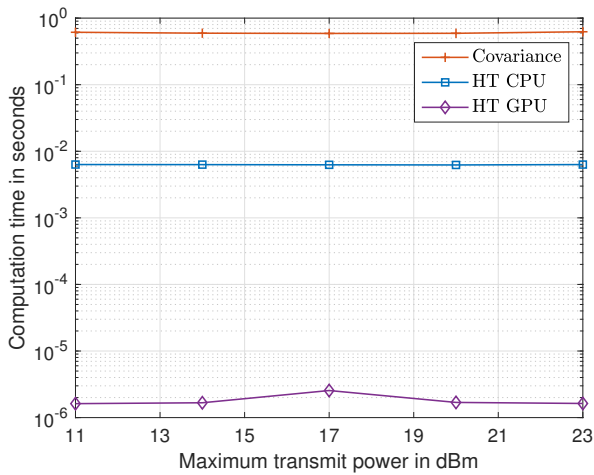}\label{T_P}
}\caption{Generalization to different SNRs.}\label{T}
\end{center}
\vspace{-0.2cm}
\end{figure}

Finally, the generalizability with respect to different SNRs is demonstrated in
Fig.~\ref{T}. The SNR
of the training samples is fixed by setting the maximum transmit power
as $p_{\max}=23$ dBm,
while the activity detection performance is tested under different
SNRs with $p_{\max}$ varying from $11$ to $23$ dBm.
The numbers of devices and BS-antennas are $N=100$ and $M=32$, respectively. As shown in Fig.~\ref{PM_P},
when the SNR decreases as the transmit power becomes lower,
the PM of different approaches becomes higher.
However, the proposed method still achieves much lower PM
than that of the covariance approach under different SNRs
for both PF = PM and PF = $2$PM.
Therefore, the proposed method generalizes well to different SNRs.
Besides the superiority of activity detection performance, Fig.~\ref{T_P}
shows that the proposed method takes significantly shorter computation time than that of the covariance approach under different SNRs.

\section{Conclusions}
This paper proposed a deep learning based method
with a customized heterogeneous transformer architecture
for device activity detection.
By adopting an attention mechanism in the neural network architecture design,
the proposed heterogeneous transformer was incorporated with desired properties
of the activity detection task. Specifically, the proposed architecture
is able to extract the relevance between device pilots and received signal,
is permutation equivariant with respect to devices, and is scale adaptable to different numbers of devices.
Simulation results showed that the proposed learning based method achieves much better activity detection performance
and takes remarkably shorter computation time than state-of-the-art covariance approach.
Moreover, the proposed method was demonstrated to generalize well to different numbers of devices, BS-antennas, and different SNRs.

\numberwithin{equation}{section}
\appendices

\section{The expression of $\text{MHA}_{\text{C}}$}\label{MHAC}
Define five sets of parameters $\mathbf{W}_t^{\text{q},\text{c}}\in\mathbb{R}^{d^{\prime}\times d}$,
$\mathbf{W}^{\text{k},\text{c}}_{\text{B},t}\in\mathbb{R}^{d^{\prime}\times d}$,
$\mathbf{W}^{\text{k},\text{c}}_{\text{Y},t}\in\mathbb{R}^{d^{\prime}\times d}$,
$\mathbf{W}^{\text{v},\text{c}}_{\text{B},t}\in\mathbb{R}^{d^{\prime}\times d}$,
and $\mathbf{W}^{\text{v},\text{c}}_{\text{Y},t}\in\mathbb{R}^{d^{\prime}\times d}$,
where $d^{\prime}$ is the dimension of each attention space and
$t\in\{1,\ldots,T\}$.
Then, we compute a query $\mathbf{q}_t^{\text{c}}$ for $\mathbf{x}_{N+1}^{[L]}$ at the $t$-th attention head:
\begin{equation}\label{qc}
\mathbf{q}_t^{\text{c}}=
\mathbf{W}_t^{\text{q},\text{c}}\mathbf{x}_{N+1}^{[L]}.
\end{equation}
The key and value corresponding to
each $\mathbf{x}_n^{[L]}$ are respectively computed as
\begin{eqnarray}\label{kvc}
\mathbf{k}_{n,t}^{\text{c}}&=&
\begin{cases}
\mathbf{W}^{\text{k},\text{c}}_{\text{B},t}\mathbf{x}_n^{[L]},~~&\forall n=1,\ldots,N,
\cr
\mathbf{W}^{\text{k},\text{c}}_{\text{Y},t}\mathbf{x}_{N+1}^{[L]},~~&n=N+1,
\end{cases}~~\\
\mathbf{v}_{n,t}^{\text{c}}&=&
\begin{cases}
\mathbf{W}^{\text{v},\text{c}}_{\text{B},t}\mathbf{x}_n^{[L]},~~&\forall n=1,\ldots,N,
\cr
\mathbf{W}^{\text{v},\text{c}}_{\text{Y},t}\mathbf{x}_{N+1}^{[L]},~~&n=N+1.
\end{cases}
\end{eqnarray}
To evaluate the relevance between $\mathbf{x}_{N+1}^{[L]}$ and each component of
$\left\{\mathbf{x}_n^{[L]}\right\}_{n=1}^{N+1}$, we
compute a compatibility $\alpha_{n,t}^{\text{c}}$ using the query $\mathbf{q}_t^{\text{c}}$
and the key $\mathbf{k}_{n,t}^{\text{c}}$:
\begin{equation}\label{compatibilityc}
\alpha_{n,t}^{\text{c}}=\frac{\left(\mathbf{q}_t^{\text{c}}\right)^T\mathbf{k}_{n,t}^{\text{c}}}{\sqrt{d^{\prime}}},~~\forall n=1,\ldots,N+1,~~\forall t=1,\ldots,T,
\end{equation}
and the corresponding attention weight is computed by normalizing $\alpha_{n,t}^{\text{c}}$ in $[0,1]$:
\begin{equation}\label{weightc}
\beta_{n,t}^{\text{c}}=\frac{e^{\alpha_{n,t}^{\text{c}}}}{\sum_{j=1}^{N+1}{e^{\alpha_{j,t}^{\text{c}}}}},~~\forall n=1,\ldots,N+1,~~\forall t=1,\ldots,T.
\end{equation}
With the attention weight $\beta_{n,t}^{\text{c}}$ scoring the relevance between $\mathbf{x}_{N+1}^{[L]}$ and each component of $\left\{\mathbf{x}_n^{[L]}\right\}_{n=1}^{N+1}$,
the attention value of $\mathbf{x}_{N+1}^{[L]}$ at the $t$-th attention head is computed as
\begin{equation}\label{attentionc}
\mathbf{x}^{\prime}_t=\sum_{n=1}^{N+1} \beta_{n,t}^{\text{c}}\mathbf{v}_{n,t}^{\text{c}}.
\end{equation}
The expression of $\text{MHA}_{\text{C}}$
is finally given by a combination of the $T$ attention values:
\begin{equation}\label{MHAcvaluec}
\text{MHA}_{\text{C}}\left(
\mathbf{x}_{N+1}^{[L]}, \left\{\mathbf{x}_n^{[L]}\right\}_{n=1}^{N}
\right)=\sum_{t=1}^T \mathbf{W}^{\text{o},\text{c}}_{t}\mathbf{x}^{\prime}_{t},
\end{equation}
where $\mathbf{W}^{\text{o},\text{c}}_{t}\in\mathbb{R}^{d\times d^{\prime}}$
is the parameter for projecting back to a $d$-dimensional vector.



\bibliographystyle{IEEEtran}
\bibliography{IEEEabrv,LiYang}

\end{document}